\documentstyle[preprint,aps]{revtex}
\begin{document}
\tightenlines
\draft
\preprint{
\parbox{4cm}{
\baselineskip=12pt
KEK-TH-581\\
TMUP-HEL-9809\\ 
\hspace*{1cm}
}}
\title{Superheavy Dark Matter and Parametric Resonance}
\author{ Masato Arai $^a$, Hideaki Hiro-Oka $^b$, 
 Nobuchika Okada $^c$ 
  \thanks{E-mail: okadan@camry.kek.jp}\thanks{JSPS Research Fellow}
     and Shin Sasaki $^a$ } 

\address{$^a$Department of Physics, Tokyo Metropolitan University,\\
         Hachioji, Tokyo 192-0397, Japan}
\address{$^b$Institute of Physics, Center for Natural Science, 
         Kitasato University\\
         Sagamihara, Kanagawa 228-8555, Japan}
\address{$^c$Theory Group, KEK, Tsukuba, Ibaraki 305-0801, Japan}

\maketitle
%
\vskip 2.5cm
\begin{center}
{\large Abstract}
\vskip 0.5cm
\begin{minipage}[t]{14cm}
\baselineskip=19pt
\hskip4mm
We propose a new scenario to produce the superheavy dark matter 
 based on the inflationary universe.  
In our scenario, the inflaton couples to both 
 a boson and a stable fermion. 
Although the fermion is produced by the inflaton decay after inflation, 
 almost energy density of the inflaton is transmitted 
 into the radiation by parametric resonance which causes 
 the explosively copious production of the boson. 
We show that the fermion produced by the inflaton decay 
 can be the superheavy dark matter, whose abundance 
 in the present universe coincides with the critical density. 
We also present two explicit models as examples
 in which our scenario can be realized. 
One is the softly broken supersymmetric theory.
The other is the ``singlet majoron model'' 
 with an assumed neutrino mass matrix.
The latter example can simultaneously explain 
 the neutrino oscillation data and the observed baryon 
 asymmetry in the present universe through the leptogenesis scenario.
\end{minipage}
\end{center}
\newpage
%
The existence of the dark matter in the present universe is 
 the commonly accepted consequence from observations \cite{kt}. 
In addition, most of the inflation models \cite{infrev}, 
 which can solve the flatness and horizon problems, 
 naturally predict the density parameter $\Omega=1$.  
On the other hand, the big-bang nucleosynthesis implies that 
 the contribution of baryons to the matter density at the present universe 
 is at most $10 \%$. 
Therefore, the present universe is dominantly fulfilled by the dark matter; 
 $\Omega_{DM} \geq 0.9$. 

In general, there are two possibilities for the type of the dark matter. 
One is that the dark matter is a thermal relic, and the other is that 
 it is a non-thermal relic. 
Most of discussions have been performed for the first type. 
In this case, the present abundance of the dark matter can 
 be estimated, and, especially, we can derive an upper bound on its mass, 
 $m_{DM} < 500$ TeV by the unitarity argument \cite{unitarity}. 
According to this argument, if the dark matter is superheavy 
 \footnote{ 
The word ``superheavy'' means that mass of the dark matter is 
 larger than 500 TeV.}, 
it should be a non-thermal relic not to over-close the present 
 universe. 
However, in this case, we should make it clear 
 what is the mechanism to produce the superheavy dark matter 
 in order to discuss its present abundance. 

Recently, some production mechanisms of the non-thermal 
 superheavy dark matter were proposed \cite{kolb}, by which 
 the dark matter is produced through gravitational effect, 
 broad parametric resonance and so on. 
Furthermore, some candidates for the superheavy dark matter were also 
 considered in the context of the string theory, 
 M-theory \cite{mtheory} and supersymmetric theory 
 with discrete gauge symmetry \cite{discrete}.  

In this letter, we propose another scenario based on 
 the inflationary universe in order to produce the superheavy dark matter 
 whose abundance in the present universe 
 coincides with the critical density. 
In our scenario, the inflaton very weakly couples to both a boson and 
 a stable fermion whose masses are much smaller than the inflaton mass. 
Although the fermion is produced by the inflaton decay after inflation, 
 almost energy density of the inflaton is rapidly transmitted 
 into the radiation of the boson by parametric resonance 
 \cite{resonance1} \cite{resonance2} 
 which causes the explosively copious production of the boson. 
If the boson couples to ordinary particles in the standard model 
 with not so weak coupling constants, the universe can be thermalized 
 as soon as the parametric resonance occurs. 
The production of the fermion is effectively over 
 when the parametric resonance occurs. 
We will show that the fermion can be the superheavy dark matter, 
 whose abundance in the present universe coincides with the critical value. 

We also present two explicit models as examples 
 in which our scenario can be realized. 
One example is the softly broken supersymmetric theory. 
Note that it is natural for a boson to simultaneously couple 
 to some bosons and fermions in the context of supersymmetric theories. 
The other is the ``singlet majoron model'' \cite{cmp} 
 with an assumed neutrino mass matrix. 
This example have other phenomenological and cosmological implications.
In this model, we can simultaneously explain the solar 
 and the atmospheric neutrino deficits and the baryon asymmetry 
 in the present universe through the leptogenesis scenario 
 \cite{leptogenesis}.    

As mentioned above, in our scenario, the inflaton field couples to 
 both the fermion and boson. 
Let us consider the inflaton potential of the form
\begin{eqnarray}
 V = \frac{1}{8} \lambda \left( \phi^2 - \sigma^2 \right)^2 \; , 
\label{potential1}
\end{eqnarray}
where $\phi$ is the inflaton field, and it is regarded 
 as a real field, for simplicity.
We also consider interaction terms of the form,
\begin{eqnarray}
 {\cal L}_{int}=  -g_B^2 \phi^2 \chi^\dagger \chi 
    - g_F \phi \overline{\psi} \psi \; , 
\label{interaction1}
\end{eqnarray}
where $\chi$ and $\psi$ are the boson and fermion fields 
 which couple to the inflaton with (positive and real) coupling 
 constants $g_B$ and $g_F$, respectively. 
Assuming $\sigma \ll m_{pl}$ , where $m_{pl}\sim 10^{19}$ GeV is 
 the Planck mass, the chaotic inflation occurs with the initial value of 
 inflaton fields, $\langle \phi \rangle \sim (\mbox{several}) 
 \times m_{pl}$. 
Note that $\lambda \sim 10^{-12}$ is implied by the anisotropy of 
 the cosmic microwave background radiation \cite{reconstruct}, 
 and $g_B$, $g_F \leq \sqrt{\lambda}$ is required by naturalness. 

Since vacuum lies at $\langle \phi \rangle = \sigma$, let us 
 shift the inflaton fields such as $\phi \rightarrow \phi+\sigma$. 
Then, we can rewrite eqs.(\ref{potential1}) and (\ref{interaction1}) as 
\begin{eqnarray}
 V = \frac{\lambda}{2} \sigma^2 \phi^2 
   + \frac{\lambda}{2} \sigma   \phi^3 
   + \frac{\lambda}{8} \phi^4  
\label{potential2}  
\end{eqnarray}
and 
\begin{eqnarray}
 {\cal L}_{int}=  -g_B^2 ( \phi^2 + 2 \sigma \phi + \sigma^2 ) 
               \chi^\dagger \chi 
     - g_F (\phi + \sigma)  \overline{\psi} \psi \; , 
\label{interaction2} 
\end{eqnarray}
respectively. 
Masses of the inflaton, boson $\chi$ and fermion $\psi$ are given by 
 $m_{\phi}= \sqrt{\lambda} \sigma$, 
 $m_{\chi}^2= g_B^2 \sigma^2 + m_0^2$ and $m_{\psi}= g_F \sigma$, 
 respectively. 
Here, $m_0$ is tree level mass of the boson field. 
In the following discussion, we assume 
 $m_{\chi}, m_{\psi} \ll m_{\phi}$. 

Now, let us consider behavior of an amplitude of the inflaton field 
 after the end of inflation. 
In this epoch, the inflaton field is coherently oscillating, 
 but its amplitude is decreasing due to 
 both the expansion of the universe and the inflaton decay. 
When the amplitude becomes $\langle \phi \rangle \ll \sigma$, 
 the inflaton potential of eq.(\ref{potential2}) is reduced to 
 only the mass term, and thus we can regard the coherent oscillation as 
 the harmonic oscillator with frequency $m_\phi$; 
 $\phi = \Phi_0 \cos (m_\phi t) $. 

Regarding the inflaton as the background field, equation of motion for 
 the Fourier modes of $\chi$ field is approximately given by
\begin{eqnarray}
  \frac{d^2}{dt^2}{\chi_k} + \left( k^2 
 +2 g_B^2 \sigma \Phi_0 
 \cos(m_\phi t)  \right) \chi_k  = 0  \; , 
\label{eom}
\end{eqnarray}
where $k$ is the spatial momentum, and we neglected mass term $m_\chi^2$ 
and $g_B^2 \phi^2$ term because of our assumption $m_\chi \ll m_\phi$ and 
 $\langle \phi \rangle \ll \sigma$, respectively. 
Here, note that we also neglected a friction term $3 H d \chi_k/ dt$ 
 introduced by the expansion of the universe, and this approximation 
 will be justified later. 
The above equation is well known as the Mathieu equation \cite{Landau} 
 , and (narrow) parametric resonance occurs if the condition 
 $4 g_B^2 \sigma \Phi_0 \ll m_\phi^2$ is satisfied.  
In the following discussion, we take $\Phi_0 \sim \sigma$ 
 and $g_B^2 \sim \lambda$ as a rough estimation. 
The amplitude of $\chi_k$ with momentum in the first resonance band
 $ k \sim m_\phi/2$, whose contribution is dominant, 
 is exponentially growing up such as $\chi_k \propto \exp(\mu t)$, where 
\begin{eqnarray}
 \mu \sim 4 \frac{g_B^2 \sigma \Phi_0}{m_\phi} \sim m_\phi \; .   
\end{eqnarray}
Note that the Hubble parameter is given by 
\begin{eqnarray} 
 H \sim \left(\frac{\lambda \sigma^2 \Phi_0^2}{m_{pl}^2}\right)^{1/2} 
 \sim \left(\frac{\sigma}{m_{pl}} \right) m_\phi \; ,
\end{eqnarray} 
and $H \ll \mu$ is satisfied because of the assumption $\sigma \ll m_{pl}$.
Then, the friction term $3 H d \chi_k / dt$ can be neglected 
 in eq.(\ref{eom}). 
Almost energy density of the inflaton field is rapidly 
 transmitted into radiation of $\chi$ fields by this parametric resonance. 
If $\chi$ field (not so weakly) couples to ordinary fields 
 in the standard model, the universe can be thermalized  
 as soon as the parametric resonance occurs. 
Then, we expect that the reheating temperature ($T_{RH}$) is estimated as 
\begin{eqnarray}
 V(\phi \sim \sigma) \sim T_{RH}^4  \; . 
 \label{reheat}
\end{eqnarray}

Next, let us consider the energy density of the fermion 
 which is produced from the time of the end of inflation ($t_{EOI}$) 
 till the time when the parametric resonance occurs 
 and the universe is thermalized ($t_{PR}$). 
At $t_{PR}$, the energy density of the produced fermion is estimated as 
\begin{eqnarray}
 \rho_{\psi}(t_{PR}) = V(t_{EOI}) 
 \left\{ 1- e^{- \Gamma (t_{PR} - t_{EOI})} \right\} 
\left( \frac{a(t_{EOI})}{a(t_{PR})}  \right)^3 \; , 
\label{fdensity}
\end{eqnarray}
where $V(t_{EOI}) \sim \lambda m_{pl}^4$ is the (potential) energy density 
 of the inflaton at $t_{EOI}$, $a(t)$ is the scale factor of the universe,  
 and $\Gamma$ is the width of the inflaton decay process 
 $\phi \rightarrow \overline{\psi} \psi$ which is given by 
\begin{eqnarray}
 \Gamma = \frac{g_F^2}{8 \pi} m_\phi \; .  
\end{eqnarray}
We regard time dependence of the scale factor 
 between $t_{EOI}$ and $t_{PR}$ as same 
 as the matter dominated universe \cite{resonance1},  
 namely, $a(t) \propto t^{2/3}$ and $H(t) \sim (2/3) t^{-1}$. 
The Hubble parameters, $H(t_{EOI})$ and $H(t_{PR})$, are given by 
\begin{eqnarray}
 H(t_{EOI}) &\sim & 
     \left(\frac{ V(\phi= m_{pl}) }{m_{pl}^2} \right)^{1/2}
   \sim  \sqrt{\lambda } m_{pl} \; , \\
  H(t_{PR}) &\sim&
     \left(\frac{V(\phi= \sigma) }{m_{pl}^2} \right)^{1/2}
   \sim  \frac{\sigma}{m_{pl}} m_\phi \; .
\end{eqnarray}
If $ \Gamma \ll H(t_{PR}) \ll H(t_{EOI})$
\footnote{
 We can see that the condition is satisfied in our final result 
 with our assumption $\sigma \ll m_{pl}$.},
 the energy density is approximately given by 
\begin{eqnarray}
\rho_{\psi}(t_{PR}) & \sim &
 \frac{2}{3} \frac{\Gamma}{H(t_{PR})} V(t_{EOI})
 \left( \frac{H(t_{PR})}{H(t_{EOI})} \right)^2  \nonumber \\
 & \sim & \frac{2}{3} \Gamma \; H(t_{PR}) \; m_{pl}^2  \; . 
\label{earlydensity}
\end{eqnarray}

The present energy density of the fermion $\psi$ is given by 
\begin{eqnarray}
 \rho_\psi(t_0) \sim \rho_\psi (t_{PR}) \times
 \left(\frac{m_\psi}{T_{RH}} \right)^4 \times
 \left(\frac{T_0}{m_\psi} \right)^3  \; ,
\label{density}
\end{eqnarray}
where $t_0$ is the present time, and $T_0 \sim 2 \times 10^{-13}$ GeV
 is the present temperature of the microwave background radiation. 
If this energy density is comparable to the critical density
 ($\rho_{crit} \sim 8 \times 10^{-47}  {\mbox{GeV}}^4)$, 
 the fermion can be the dark matter. 
Using eqs.(\ref{reheat})-(\ref{density}), 
 we can get resultant mass ratio of the dark matter to the inflaton 
\footnote{
 We get the small coupling constant $g_F \sim 10^{-9}$ 
 from eq.(\ref{result}). 
 It is easy to check that this coupling constant is too small 
 for the superheavy dark matter to be in thermal equilibrium 
 with other fields. 
 This fact is consistent with our assumption.},  
\begin{eqnarray}
 m_\psi / m_\phi  \sim 10^{-3} \; .  
\label{result}
\end{eqnarray}
If we take, for example, $\sigma \sim 10^{18}$ GeV, the fermion 
 can be the superheavy dark matter with mass $m_\psi \sim 10^9$ GeV.  

Note that the existence of the parametric resonance is crucial 
 in our scenario. 
If there is no parametric resonance, 
 according to the old scenario of reheating, 
 the reheating temperature is estimated as 
\begin{eqnarray}
 \Gamma = H \sim \frac{T_{RH}{}^2}{m_{pl}} \; , 
\end{eqnarray}
where $\Gamma$ is the decay width of the inflaton. 
However, in this scenario, the energy density of the fermion 
 in the early universe is comparable with that of radiation, 
 and thus the present universe is over-closed,  
 if not $g_F$ is unnaturally small compared with $g_B$. 

Here, we give a comment on our analysis in this letter. 
In general, there is a possibility that the broad resonance occurs  
 before the amplitude of the inflaton is dumped into the narrow 
 resonance band. 
However, it is non-trivial problem to definitely decide 
 whether the broad resonance can affect or not, 
 because of its stochastic behavior as analyzed in \cite{resonance2}. 
Then, in our analysis we ignore the effect of the broad resonance 
 for simplicity. 
If the energy of the inflaton is transmitted enough 
 by the broad resonance, the period of reheating is shorten 
 and total number of $\psi$ produced by the inflaton decay is 
 smaller than that of our estimation. 
In this case, the resultant mass of the superheavy dark matter 
 becomes larger than that of our estimation 
 in order for its present abundance to coincide with the critical value. 
It is always possible to understand our result as the lower bound 
 on the super heavy dark matter mass. 
Some numerical calculations are needed for correct evaluation. 

In the following, we discuss two explicit models as examples 
 in which our scenario of the superheavy dark matter production 
 can be realized.    
The first example is the softly broken supersymmetric theory. 
Let us consider superpotential of the form
\begin{eqnarray}
 W = \frac{1}{2} m_Z Z^2 + \frac{1}{2}m_X X^2 + \lambda_Z Z X^2 \; ,
 \label{superpotential}
\end{eqnarray}
where $Z$ and $X$ are superfields with masses $m_Z$ and $m_X$, respectively, 
 and $\lambda_Z$ is the dimension-less coupling constant. 
In the following discussion, 
 we assume $m_X \ll m_Z$ and $\lambda_Z \ll 1$, and 
 identify the scalar component of $Z$ 
 and fermion component of $X$ ($\psi_X$) with the inflaton 
 and the dark matter 
\footnote{
 The fermion can be stable, if we introduce a discrete symmetry
 such as R-parity.}, respectively. 
The scalar potential is given by
\begin{eqnarray}
 V&=& m_Z^2  Z^\dagger Z + m_X^2 X^\dagger X  + \lambda_Z^2 (X^\dagger X)^2 
    + 4 \lambda_Z^2 (Z^\dagger Z) (X^\dagger X) \nonumber \\
  &+& 
  \left\{ \lambda_Z m_Z Z X^{\dagger 2} 
   + 2 \lambda_Z m_X Z (X^\dagger X) + \mbox{h.c.}  \right\} \; .
\label{scalarpotential}
\end{eqnarray}
Note that $m_Z \sim 10^{13}$ GeV is required by the anisotropy 
 of the microwave background radiation \cite{reconstruct}. 

The inflaton field $Z$
\footnote{
 We use the same notation both superfields and their scalar 
 components in this letter.}
 is coherently oscillating after the end of inflation 
 and this oscillation is just the harmonic oscillator with 
 frequency $m_Z$; $Z= \Phi_0 \sin(m_Z t)$. 
In the following, we consider the case in which the fourth term, 
 $4 \lambda_Z^2 (Z^\dagger Z) (X^\dagger X)$,  
 in eq.(\ref{scalarpotential}) is dominant compared with the second line. 
For example, this case is realized, if we introduce 
 the soft supersymmetry breaking terms which cancel the terms 
 in the second line. 
Regarding the inflaton as the background field, the equation of motion 
 for the Fourier mode of the scalar $X$ is given by 
\begin{eqnarray}
 X^{\prime \prime}_k + \left( A_k - 2 q \cos 2z \right) X_k =0 \; ,
\label{mathiew} 
\end{eqnarray} 
where $A_k= k^2/m_Z^2 + 2 q$, $q = (\lambda_Z \Phi_0 / m_Z)^2$, $z=mt$, 
 and the prime denotes differential with respect to $z$. 
This is nothing but the Mathieu equation, 
 and the (narrow) parametric resonance occurs if we fix parameters 
 as $\lambda_Z \ll 1$ and $q \ll 1$, 
 which means $\lambda_Z \ll 10^{-6}$ 
 \footnote{
 This coupling constant is also too small for the superheavy dark matter 
 to be in thermal equilibrium with other fields.}
(we take $\Phi_0 \sim m_{pl}$, see the following discussion). 

Since the oscillating inflaton is regarded as the harmonic oscillator 
 from the beginning, the resonance occurs just after the end of inflation. 
Thus, it is naturally expected that $H(t_{EOI}) \sim H(t_{PR})$ and 
 the reheating temperature is given by 
 $T_{RH}^4 \sim m_Z^2 \Phi^2_0 \sim m_Z^2 m_{pl}^2$. 
Following the discussion from eq.(\ref{fdensity}) to eq.(\ref{density}), 
 the present energy density of the fermion component of $X$ is described as
\begin{eqnarray}
 \rho_{\psi_X} \sim \frac{2}{3} \Gamma \; m_{pl}^2 \; m_Z  \times 
  \left( \frac{m_{\psi_X}}{T_{RH}} \right)^4 \times 
  \left( \frac{T_0}{m_{\psi_X}} \right)^3  \sim \rho_{crit} \; ,
\end{eqnarray}
where $\Gamma= \lambda_Z^2 m_Z /8 \pi$ is the inflaton decay width 
 with respect to the channel $Z \rightarrow \psi_X \psi_X$. 
We can get the mass of the dark matter as 
 $m_{\psi_X} \sim 10^{-7}/ \lambda_Z^2 \gg 10^5$ GeV. 
This result depends on the parameter $\lambda_Z$. 

In the above discussion we assumed that the universe can 
 be thermalized as soon as the parametric resonance occurs.  
This can be possible if we introduce the soft supersymmetry breaking term 
 (A-term) such as 
\begin{eqnarray}
 {\cal L}_{soft} = A X {\bar H} H \; ,
\end{eqnarray}
where $A$ is a parameter of mass dimension one, and 
 ${\bar H}$ and $H$ is the down- and up-type Higgs doublets 
 in the supersymmetric standard model, respectively. 
Since the scalar $X$ couples to the standard model particles 
 by this soft terms, the thermalization of the universe is realized. 
  
Next, let us discuss the second example. 
There are some current data suggesting masses and flavor mixing 
 of neutrinos. 
The solar neutrino deficit \cite{solar} and the atmospheric neutrino 
 anomaly \cite{atm} seem to be indirect evidences for the neutrino mass 
 and flavor mixing from the viewpoint of neutrino oscillation. 
The ``singlet majoron model'' \cite{cmp} is a simple extension 
 of the standard model to give Majorana masses to neutrinos. 
It is well known that this model includes the see-saw mechanism 
 \cite{seesaw}, which can naturally explain the smallness of 
 the neutrino masses compared with other leptons and quarks. 

In addition, the model has a cosmological implication. 
Since, in the model, the lepton number is broken 
 and CP is also violated by non-zero CP-phases in general, 
 we can explain the observed baryon asymmetry in the present universe 
 through the leptogenesis scenario \cite{leptogenesis}. 

Assuming a typical matrix for the neutrino mass matrix, 
 the singlet majoron model can be an example 
 to which our scenario can apply. 
We can also simultaneously explain the observations of the solar and 
 atmospheric neutrino deficits and the baryon asymmetry 
 in the universe through the leptogenesis scenario. 

We introduce the Yukawa interaction of the form 
\begin{eqnarray}
 {\cal L}_Y = - g_{{}_Y}^{ij} \overline{\nu_L}^i \phi_d \nu_{R }^j 
 -g_{{}_M }^{kl} \overline{\nu_{R}{}^c}^k \phi_s \nu_{R}^l  +h.c. \; ,
 \label{yukawa}
\end{eqnarray}
where $\phi_d$ is the neutral component of the Higgs doublet 
 in the standard model, $\phi_s$ is the electroweak singlet 
 Higgs field, and $i,j,k,l=1,2,3$ are generation indices. 
The Dirac and Majorana mass terms appear by non-zero vacuum expectation 
 values of these Higgs fields. 
The neutrino mass matrix is given by 
\begin{eqnarray}
 \left[ \begin{array}{cc}
   0  &  m_{{}_D} \\  m_{{}_D}^T & M 
 \end{array}\right]
 \label{mat}\; ,
\end{eqnarray}
where $m_{{}_D}=g_{{}_Y}^{ij}\langle \phi_d \rangle$ is the Dirac mass term, 
 and $M=g_{{}_M}^{kl}\langle \phi_s \rangle$ is the Majorana mass term. 

We assume typical Yukawa coupling constants and mass matrices as follows: 
\begin{eqnarray}
m_{{}_D}=  \left[ \begin{array}{ccc}
 0 & ~~a  & 0 \\ 
 0 & ~~b  & 0 \\
 0 & ~~b  & ~b e^{ i \delta}   
 \end{array}\right]  \label{mat2}\; ,  \; \; 
M=  \left[ \begin{array}{ccc}
 M_1  &  0   & 0  \\
  0   &  M_2 & 0  \\
  0   &  0   & M_3  \end{array}\right]  \; . 
\label{matrix}
\end{eqnarray}
Here, $a$, $b$ and $\delta$ are real parameters, and 
 $M_1$, $M_2$ and $M_3$ are real masses of the heavy Majorana neutrinos. 
Note that, since the right-handed neutrino of the first generation
 decouples to the left-handed neutrinos, 
 it can be stable if its mass is smaller than the singlet Higgs mass. 
This is the case we consider. 

In this model, we can identify the inflaton and the boson $\chi$ 
 in eqs.(\ref{potential1}) and (\ref{interaction1})
 with the singlet Higgs $\phi_s$ and 
 the standard model Higgs doublet, respectively. 
Then, our scenario can be realized, and thus the right-handed neutrino 
 of the first generation can be the superheavy dark matter. 
The mass of the dark matter is three orders of magnitude smaller 
 than the singlet Higgs mass, $M_1 \sim 10^{-3} \times m_{\phi_s}$, 
 according to our result of eq.(\ref{result}). 

Next, we consider the problems of the solar 
 and the atmospheric neutrino deficits.
By the see-saw mechanism, the mass matrix of eq.(\ref{mat}) 
 is approximately diagonalized as 
\begin{eqnarray}
\left[ \begin{array}{cc}
  m_{{}_D} M^{-1} m_{{}_D}^T  & 0  \\
  0   &  M  
 \end{array}\right]  
 \label{diag1}\; ,
\end{eqnarray}
where the mass matrix for the light neutrinos, 
 $m_{{}_D} M^{-1} m_{{}_D}^T$, is given by 
\begin{eqnarray}
 m_{{}_D} M^{-1} m_{{}_D}^T = 
\left[ \begin{array}{ccc}
  a^2 / M_2  & ~a b / M_2  &  a b / M_2 \\
  a b / M_2  & ~b^2 / M_2  &  b^2 / M_2  \\ 
  a b / M_2  & ~b^2 / M_2  & ~b^2 / M_2 +b^2 e^{2 i\delta}/M_3
 \end{array}\right]  \; . 
\label{diag2}
\end{eqnarray}
Note that $M_1$ is absent in this matrix
 since the right-handed neutrino is decoupled to
 left-handed neutrinos.
For simplicity, we assume $a/b \equiv \epsilon \ll 1$, 
 $M_2 /2 M_3  \equiv \gamma \ll 1$ and $\delta \ll 1$. 
Neglecting the elements higher than the second order 
 with respect to $\epsilon $, $\gamma$ and $\delta$, we can get 
\begin{eqnarray}
 m_{{}_D} M^{-1} m_{{}_D}^T \sim  \frac{b^2}{M_2}
 \left[ \begin{array}{ccc}
     0      & ~~\epsilon & \epsilon \\
  \epsilon  & ~~   1     &     1    \\
  \epsilon  &  ~~  1     &  1 + 2 \gamma 
 \end{array}\right]  \; .
\label{diag3}
\end{eqnarray}

The unitary matrix which diagonalizes this matrix is the physical 
 Kobayashi-Maskawa matrix in the lepton sector at low energies. 
Moreover, assuming $ \epsilon \ll \gamma $, 
 we can get hierarchical mass eigenvalues such as
 $b^2/M_2 (0, \; \gamma, \; 2 + \gamma)$ by diagonalizing this matrix. 
The mass squared differences and mixing angles corresponding  
 to the solar and atmospheric neutrino oscillation data are given by 
\begin{eqnarray}
& & \Delta m^2_{\odot} \sim  \left(\frac{b^2}{M_2}\right)^2 \gamma^2 
 \; \;  ; \; \;    
 \sin^2 2 \theta_\odot \sim 4 
 \left(\frac{\epsilon}{\gamma} \right)^2  \ll 1 \; , \nonumber \\
& & \Delta m^2_\oplus \sim 4 \left(\frac{b^2}{M_2}\right)^2 
 \; \;  ; \; \;
 \sin^2 2 \theta_\oplus \sim 1  \; . 
\label{oscildata}
\end{eqnarray}
By the assumption $\epsilon \ll \gamma$, 
 our description for the solar neutrino data corresponds to 
 the small angle MSW solution \cite{msw}. 
By appropriately choosing the values of the free parameters, 
 we can reproduce the solar neutrino data 
 $\Delta m_\odot^2 \sim 5 \times 10^{-6} \mbox{eV}^2$
 for the small angle MSW solution \cite{hata} 
 and the atmospheric neutrino data 
 $\Delta m_\oplus^2 \sim 5 \times 10^{-3} \mbox{eV}^2$ \cite{atm}. 

On the other hand, we can also explain the baryon asymmetry 
 in the present universe through the leptogenesis scenario. 
Since the right-handed neutrinos in the second and the third generations 
 couple to the left-handed neutrinos and $m_{{}_D}$ of eq.(\ref{matrix})
 includes CP-phase $\delta$, 
 the original scenario of the leptogenesis can work. 
According to the original work \cite{leptogenesis}, 
 we can calculate the net lepton number production rate, 
 which is proportional to 
\begin{eqnarray}
 \Delta  = \frac{1}{\pi v^2 (m_{{}_D}^\dagger m_{{}_D})_{22}} \;  
 \mbox{Im} \left[ 
 (m_{{}_D}^\dagger m_{{}_D})_{23}
 (m_{{}_D}^\dagger m_{{}_D})^T_{32} \right]f(M_3^2/M_2^2)  \; .  
\end{eqnarray}
Here, $f(M_3^2/M_2^2) \sim M_2/ 2 M_3 = \gamma $ for $M_2 \ll M_3$, 
 and $v=246$ GeV is the vacuum expectation value of the standard model 
 Higgs doublet. 
Using the Dirac mass matrix of eq.(24) and eq.(\ref{oscildata}),  
 $\Delta$ is given by 
\begin{eqnarray}
 \Delta \sim  \frac{ \sqrt{ \Delta m^2_\odot} M_2} {\pi v^2} \delta \; .
\end{eqnarray} 
In the following, to consider the observed baryon to photon ratio 
 $n_B/n_\gamma \sim 10^{-10}$, 
 we use the rough estimation $n_B/n_\gamma \sim \Delta/g_*$ \cite{kt}, 
 where $g_* \sim 100$ is the effective degrees of freedom 
 of the radiation in the early universe.  

Choosing appropriate values for the parameters $M_2$ and $\delta$, 
 we can simultaneously explain the correct abundance 
 of the superheavy dark matter, neutrino oscillation data 
 and the observed baryon ratio.  
When we take $\delta = 0.1$, for example, 
 we can get the mass spectrum as follows:  
\begin{eqnarray}
 M_1 \sim 10^9 \; < \; M_2 \sim 10^{10} \; < \;  
 M_3 \sim 10^{11} \; < \; m_{\phi_s} \sim 10^{12} \; \mbox{GeV} \;.  
\end{eqnarray}
Here, we took the vacuum expectation value of the singlet Higgs 
 as $\sigma = 10^{18}$ GeV. 

In summary, we proposed a new scenario to produce the superheavy dark matter.
In our scenario, the inflaton couples to both the boson 
 and the stable fermion whose masses are much smaller than the inflaton mass.
The fermion is produced by the decay of inflaton after the end of inflation, 
 but this production is effectively over when the parametric 
 resonance occurs and 
 the copious bosons are explosively produced. 
We showed that the abundance of the fermion in the present universe 
 can be comparable to the critical value, and the fermion can be 
 the superheavy dark matter. 
In addition, we presented two explicit models as examples  
 in which our scenario can be realized.
One is the softly broken supersymmetric theory.  
In this example, the dark matter has mass of larger than $10^5$ GeV. 
The other example is the ``singlet majoron model'' 
 with the assumed neutrino mass matrix. 
In this model, the right-handed neutrino of the first generation 
 can be the superheavy dark matter with mass, for example, $10^9$ GeV.
Furthermore, this model can simultaneously explain 
 the neutrino oscillation data and the observed baryon asymmetry 
 in the universe through the leptogenesis scenario. 

One of the authors (N.O.) would like to thank Noriaki Kitazawa 
for useful discussion and comments, and also for his hospitality 
 during the author's visit to Yale University 
 where some parts of this work were completed. 
This work was supported in part by a Grant-in-Aid for Scientific Research 
 from the Ministry of Education, Science and Culture and Research Fellowship 
 of Japan Society for the Promotion of Science for Young Scientists. 

%
\end{document}